\documentclass[12pt]{article}
\usepackage{lscape,slashed}
\usepackage{multirow}
\usepackage{color}
\usepackage{bm}
\usepackage{comment}
\usepackage{amssymb,amsmath,yhmath,mathrsfs,graphicx}
\usepackage{amstext}
\usepackage{amscd}
\usepackage{graphics}
\usepackage{epsfig}
\usepackage{shadow}
\usepackage[all]{xy}
\usepackage{hyperref}

\def\be{\begin{equation}}
\def\ee{\end{equation}}
\def\bes{\begin{equation*}}
\def\ees{\end{equation*}}
\def\beq{\begin{equation}}
\def\eeq{\end{equation}}
\def\bea{\begin{eqnarray}}
\def\eea{\end{eqnarray}}
\def\beas{\begin{eqnarray*}}
\def\eeas{\end{eqnarray*}}

\def\sideremark#1{\ifvmode\leavevmode\fi\vadjust{\vbox to0pt{\vss
 \hbox to 0pt{\hskip\hsize\hskip1em
 \vbox{\hsize2cm\tiny\raggedright\pretolerance10000
  \noindent #1\hfill}\hss}\vbox to8pt{\vfil}\vss}}}

\begin{document}
\thispagestyle{empty}

\setcounter{footnote}{0}

\vspace{-50mm}
\begin{flushright}\tt CALT-TH 2015-055,\! BRX-TH-6300 \end{flushright}

\begin{center}

{\Large
 {\bf  Non-linear duality invariant partially massless models?}\\[4mm]

 {\sc \small
  D.~Cherney$^{\rm b}$,   S.~Deser$^{\rm a}$, A.~Waldron$^{\rm b}$ and G. Zahariade$^{\rm c}$}\\[3mm]

{\em\small
                      ~\ ${}^\mathfrak{\rm a}$
                       Walter Burke Institute for Theoretical Physics, California Institute of Technology, Pasadena, CA 91125; 
     Physics Department, Brandeis University, Waltham, MA 02454.\\
{\tt deser@brandeis.edu}\\[2mm]
          
           ~${}^{\rm b}\!$
            Department of Mathematics\
            University of California,
            Davis CA 95616, USA\\
            {\tt wally@math.ucdavis.edu}

           ~${}^{\rm c}\!$
            Department of Physics\
            University of California,
            Davis CA 95616, USA\\
            {\tt  zahariad@ucdavis.edu}\\[2mm]

            }
}

\bigskip

{\sc Abstract}\\[1mm]

\end{center}

\newcommand{\mGR}{$\overline {\text m\hspace{-.1mm}}$GR\ }

\noindent

We present manifestly duality invariant, non-linear, equations of motion for maximal depth, partially massless higher spins.  These are based on a first order, Maxwell-like formulation of the known partially massless systems. Our models mimic  Dirac-Born-Infeld theory but it is unclear whether they are Lagrangian.

\newpage



\section{Introduction}

In four-dimensional de Sitter (dS) space there exist novel ``photon-like'' excitations---the maximal depth, spin~$s$, partially massless (PM) theories~\cite{PM}.
These propagate lightlike helicities $(\pm s, \pm (s-1),\ldots \pm 1)$, the zero helicity state being removed by a scalar gauge invariance~\cite{PMc}. Viewing this as a $U(1)$ invariance, 
these models can be coupled to charged matter~\cite{PMEM}.
Moreover, these (linear) models are conformally invariant~\cite{DWso}, enjoy a Maxwell-like duality invariance~\cite{DWdu} and  have monopole solutions~\cite{Hintrose}. This duality was first demonstrated as a symmetry of the model's actions in a Hamiltonian formulation in~\cite{DWdu}. Subsequently, a manifestly covariant proof of this duality was given~\cite{Hintman} for the spin~2 PM system at the level of the equations of motion: 
\begin{equation}\label{dFdeltaF}
\nabla^\mu F_{\mu\nu\rho} = 0 = \nabla_{[\mu} F_{\nu\rho]\sigma}\, ;\quad F_{\mu\nu\rho}=-F_{\nu\mu\rho}\, .
\end{equation}
These can be shown to be equivalent to the standard PM equations of motion for a   symmetric, rank~2,  potential $A_{\mu \nu}$ where
$F_{\mu\nu\rho}= 2\nabla_{[\mu} A_{\nu]\rho}$
and  $\nabla$ is the Levi-Civita connection of the background dS metric. The curvature $F_{\mu\nu\rho}$ enjoys the scalar PM gauge invariance
$$
A_{\mu\nu}\sim  A_{\mu\nu}+\big(\nabla_\mu \nabla_\nu+\frac\Lambda 3 g_{\mu\nu}\big) \alpha\, .
$$
The equations of motion~\eqref{dFdeltaF} are manifestly invariant under the interchange $F_{\mu\nu\rho}\leftrightarrow F_{\widetilde{\mu\nu}\rho}$ where \scalebox{1.1}{$\widetilde{\scriptstyle \cdots}$} denotes the Hodge~$\star$ operation; indeed, these linear models enjoy continuous duality invariance in terms of their canonical variables~\cite{DWdu}.

Our aim is to search for a non-linear generalization of
these models. For the spin~1, Maxwell ancestor of Equation~\eqref{dFdeltaF},  such a generalization has been long known---the Dirac--Born--Infeld (DBI) theory~\cite{DBI}. 
[Note also that (non-linear) conformal/Weyl gravity enjoys duality under discrete interchange of electric and magnetic curvatures~\cite{DN1}. About its flat or deSitter vacua, it propagates
 both graviton and PM modes\footnote{One might speculate that integrating out the graviton excitations from a conformal gravity
path integral could lead to a duality invariant, non-linear PM model. Note however, that already classically it is not possible to truncate conformal gravity to a non-linear PM sector~\cite{DJW}. Also these excitations are (necessarily) relatively ghost.}~\cite{Maldacena}.] 
Succinctly, our aim is to construct  ``partially massless DBI'' (PM-DBI) models.

We follow the treatment of Maxwell's equations in a medium
\begin{equation}\label{dGdeltaF}
\nabla^\mu G(F)_{\mu\nu}=0=\nabla_{[\mu}F_{\nu\rho]}\, ,
\end{equation}
given  in terms of electromagnetic fields $F_{\mu\nu}$ and their accompanying electric intensity and magnetic inductions described by some non-linear function $G(F)^{\mu\nu}$. In particular we show (following earlier electromagnetic DBI analyses of~\cite{GR}) how to construct the analogous higher spin constitutive relations $G(F)$ such that Equation~\eqref{dGdeltaF} and its $s\geq 2$ counterparts  are duality invariant.

\section{First-order formulation}
\label{fof}

The totally symmetric, rank~$s$ potentials~$A_{\mu_1\ldots \mu_s}$ of the maximal depth PM spin~$s$ systems
are defined up to  order~$s$ in derivatives gauge transformations.
Their gauge invariant curvatures 
are then given as first derivatives of the potentials. The advantage of working with curvatures instead of potentials as the basic dynamical variables is that we need not concern ourselves with gauge invariance.
We thus consider two-form, trace-free symmetric tensor-valued curvatures~$F_{\mu\nu\alpha_1\ldots \alpha_{s-1}}$, so that
$$
F_{\mu\nu\alpha_1\ldots \alpha_{s-1}}=
-F_{\nu\mu\alpha_1\ldots \alpha_{s-1}}=
F_{\mu\nu(\alpha_1\ldots \alpha_{s-1})}\, ,\quad
F_{\mu\nu}{}^\alpha{}_{\alpha\alpha_3\ldots \alpha_{s-1}}=0\, .
$$
We then impose equations of motion analogous to
the spin~2 PM system~\eqref{dFdeltaF}
\begin{equation}\label{dFdeltaFs}
\nabla^\mu F_{\mu\nu\alpha_1\ldots \alpha_{s-1}}=0=\nabla_{[\mu}F_{\nu\rho]\alpha_1\ldots \alpha_{s-1}}\, .
\end{equation}
Conjecturally, these equations describe the maximal depth PM system for any spin $s$. 
Spin~$1$ is of course just the dS Maxwell system, while this statement was proven for spin~2 in~\cite{Hintman} (based on earlier works~\cite{PMEM,SV}). We have explicitly verified that for $s=3$ these equations describe maximal depth PM\footnote{The $s\geq 4$ models remain, therefore, conjectural. However, Kurt Hinterbichler has informed us that he and collaborators have a general construction of first order PM equations of motion including  also  spins~$s$~$\geq$~$4$.}. To see this, we need to verify that the above equations propagate six electric and six magnetic degrees of freedom with helicities 
$(\pm 3,\pm2,\pm 1)$. To begin with there are $54$ dynamical curvature fields subject to the $72$ equations of motion in~\eqref{dFdeltaFs}. Specializing to Hubble coordinates $(t,x^i)$ with metric
$$
ds^2= -dt^2 + e^{2\sqrt{\Lambda/3}\, t} (dx^2+dy^2+dz^2)\, ,
$$
we see that there are $18$ primary constraints (devoid of time derivatives) on dynamical fields:
\begin{equation*}
\nabla^\mu F_{\mu t\alpha\beta}=0=\nabla_{[i}F_{jk]\alpha\beta}\, .
\end{equation*}
Taking a further covariant divergence or curl of the equations of motion~\eqref{dFdeltaFs} and using that the dS space has constant curvature implies that
$$
F_{(\alpha|\nu}{}^\nu{}_{|\beta)}=0=
\varepsilon_{(\alpha|}{}^{\mu\nu\rho}
F_{\mu\nu\rho|\beta)}\, .
$$
These two relations are identically trace-free and thus impose $18$ secondary constraints. Finally we must find six further 
tertiary constraints and verify that only helicities $(\pm 3,\pm2,\pm 1)$ propagate. For that one can Fourier transform over  the three spatial coordinates $x^i$, so that $\partial_i=i k_i$ and then explicitly solve both the primary and secondary constraints. 
Choosing, without loss of generality, 
$k_i=(0,0,1)$, it is then easy, but tedious, to verify that the remaining equations of motion determine the evolution of
six electric 
$
F_{t(abc)_\circ}, F_{t(ab)_\circ z}, F_{tazz}
$
and magnetic
$
F_{z(abc)_\circ}, F_{z(ab)_\circ z}, F_{zazz}
$ fields
($a,b,c = 1,2$ and $(\cdots)_\circ$ denotes trace-free symmetrization) with respective helicities $(\pm3, \pm2,\pm 1)$.

We now consider possible non-linear generalizations of the Equations~\eqref{dFdeltaFs} along the lines of 
Maxwell's equations in a medium \begin{equation}\label{dGdeltaFs}
\nabla^\mu G(F)_{\mu\nu\alpha_1\ldots \alpha_{s-1}}=0=\nabla_{[\mu}F_{\nu\rho]\alpha_1\ldots \alpha_{s-1}}\, ,
\end{equation}
for some invertible, derivative-free, functional $G(F)$ with the same symmetries as the curvatures~$F$.
For the Maxwell system, this maneuver does not alter the propagating degree of freedom count
so long as $G(F)$ is chosen such that the two equations above are independent.
For the higher spin $s\geq 3$ PM systems this is no longer obvious,
although at least the primary and secondary constraints 
required for  a correct degree of freedom count follow from the  argument outlined above for~$s=3$. We have not studied 
what requirements   tertiary and higher constraints place on the functional~$G(F)$ for spins $s\geq 3$. Hence currently we only have a proof for spin~2 that equations~\eqref{dGdeltaFs} propagate the correct degrees of freedom.

\section{Duality}

Suppose one is given a space of two-form curvatures $\{{\mathcal F}\}$ and an infinitesimal symmetry transformation
$$
\delta {\mathcal F}=  \star \, G({\mathcal F})
$$
with the involutive property
\begin{equation}\label{involute}
\delta \big(G({\mathcal F})) \propto \star\,  {\mathcal F}\, .
\end{equation}
Then the system of equations
$$
\left\{
\begin{array}{l}
{\mathcal B}({\mathcal F})=0\\[2mm]
{\mathcal B}(\star G({\mathcal F})) = 0\, ,
\end{array}\right.
$$
is manifestly invariant under the symmetry~$\delta$ for any linear functional ${\mathcal B}$.
The duality invariant Maxwell system is obtained this way by taking ${\mathcal B}$
to be the covariant curl. Then $G$ is the identity map and $\delta$ is the standard electromagnetic duality symmetry.
More general electromagnetic solutions to the involutive requirement~\eqref{involute} can be obtained by a generating functional  ansatz~\cite{GR} (see also~\cite{PS})
\begin{equation}
\label{EMcon}
G(F)^{\mu\nu}=-\frac{2}{\sqrt{-g}}\, \frac{\delta S(F)}{\delta F_{\mu\nu}}\, .
\end{equation}
Imposing equation~\eqref{involute}
then implies~\cite{GR}
\begin{equation}\label{EBDH}
F^{\mu\nu} F_{\widetilde{\mu\nu}}=
 G(F)^{\mu\nu} G(F)_{\widetilde{\mu\nu}}
+ {\rm constant}\, .
\end{equation}
Notice that the Maxwell action $S(F)=-\frac14 \int d^4x  \sqrt{-g}\,  F_{\mu\nu} F^{\mu\nu}$ 
gives $G(F)^{\mu\nu}=F^{\mu\nu}$
and thus satisfies the above requirement.  The only other solution to Equation~\eqref{involute} based on the above ansatz whose linearized dynamics recovers Maxwellian electromagnetism, is the DBI action~\cite{DBI,Attaturk} 
\begin{equation}\label{DBI}
S(F)=-\mu^4\int d^4x \sqrt{-g}\, \sqrt{1+\frac1{2\mu^4} F_{\mu\nu}F^{\mu\nu}-\frac 1{16\mu^8} \big(F_{\widetilde{\mu\nu}}F^{\mu\nu} \big)^2}\, .
\end{equation}
Here the $\mu$ is a parameter with dimensions of mass which we henceforth set to unity.
We are now ready to investigate whether this duality mechanism extends to higher spins.

\section{PM Duality}

 For higher spin PM systems, we thus make an ansatz analogous to~\eqref{EMcon} for the constitutive relations
 $$
 G(F)^{\mu\nu\alpha_1\ldots \alpha_{s-1}}
 =-\frac{2}{\sqrt{-g}}\, \frac{\delta S(F)}{\delta F_{\mu\nu\alpha_1\ldots \alpha_{s-1}}}\, .
 $$
 The involutive requirement~\eqref{involute} now imposes
 \begin{equation}\label{EBDHs}
F^{\mu\nu\alpha_1\ldots \alpha_{s-1}} F_{\widetilde{\mu\nu}\alpha_1\ldots \alpha_{s-1}}=
 G(F)^{\mu\nu\alpha_1\ldots \alpha_{s-1}} G(F)_{\widetilde{\mu\nu}\alpha_1\ldots \alpha_{s-1}}
+ {\rm constant}\, .
\end{equation}
To solve this equation one should find a basis for all possible (covariant) scalars built from curvatures. When $s=1$, there are only two possibilities $F_{\mu\nu}F^{\mu\nu}$ and $F_{\widetilde{\mu\nu}}F^{\mu\nu}$.
This allows Equation~\eqref{EBDH} to be reformulated as the problem of finding an exact  unit vector on a Riemannian two-manifold coordinatized  by these two variables~\cite{GR}. For the case $s=2$,  we present in Appendix~\ref{6manifold}  an analogous 6-manifold version of this problem obtained by expressing $S(F)$ in terms of curvature bilinears. [Generally, for $s\geq 2$, one can also consider scalars built from higher powers of curvatures.] Our present aim is not to map out a space of all possible non-linear duality-invariant models, but instead to study the simplest
 of these, directly inspired by the DBI functional~\eqref{DBI}.

Consider now the functional
\begin{equation}\label{bigformaggio}
S(F)= \int d^4x \sqrt{-g}\,{\mathcal L} \, ,
\end{equation}
where 
$$
{\mathcal L}:=- \sqrt{1+\frac1{2} F_{\mu\nu\alpha_1\ldots \alpha_{s-1}}F^{\mu\nu\alpha_1\ldots \alpha_{s-1}}-\frac 1{16} \big(F_{\widetilde{\mu\nu}\alpha_1\ldots \alpha_{s-1}}F^{\mu\nu\alpha_1\ldots \alpha_{s-1}} \big)^2}\, .
$$
This gives the higher spin constitutive relations\footnote{These relations can be inverted:
$$
F(G)_{\mu\nu\alpha_1\ldots \alpha_{s-1}}=-\, {\mathcal K}^{-1}\, \Big(G_{\mu\nu\alpha_1\ldots \alpha_{s-1}}
+{\frac{\textstyle1}{\, \textstyle4\, }} \, \big(G_{\widetilde{\rho\sigma}\beta_1\ldots \beta_{s-1}}G^{\rho\sigma\beta_1\ldots \beta_{s-1}}\big)\,G_{\widetilde{\mu\nu}\alpha_1\ldots \alpha_{s-1}} \Big)\, ,
$$ 
where 
$$
{\mathcal K}:=\sqrt{1-\frac1{2} G_{\mu\nu\alpha_1\ldots \alpha_{s-1}}G^{\mu\nu\alpha_1\ldots \alpha_{s-1}}-\frac 1{16} \big(G_{\widetilde{\mu\nu}\alpha_1\ldots \alpha_{s-1}}G^{\mu\nu\alpha_1\ldots \alpha_{s-1}} \big)^2}\, .
$$}
\begin{equation}\label{sconst}
G(F)_{\mu\nu\alpha_1\ldots \alpha_{s-1}}=-\, \frac{F_{\mu\nu\alpha_1\ldots \alpha_{s-1}}
-{\frac{\textstyle1}{\, \textstyle4\, }} \, \big(F_{\widetilde{\rho\sigma}\beta_1\ldots \beta_{s-1}}F^{\rho\sigma\beta_1\ldots \beta_{s-1}}\big)\,F_{\widetilde{\mu\nu}\alpha_1\ldots \alpha_{s-1}} }{\mathcal L}\, .
\end{equation}
It is easy to verify that these obey Equation~\eqref{EBDHs} with zero constant term (as  for the electromagnetic DBI theory). Since the 
the map $F\mapsto G(F)$ given in~\eqref{sconst} is invertible, 
as discussed in Section~\ref{fof},  
at least for $s=2$ the equations of motion~\eqref{dGdeltaFs} propagate the correct degree of freedom count. 
Therefore, for $s=2$, the above constitutive relation defines non-linear, dS covariant, 
duality invariant PM equations of motion. 
As explained earlier, to prove the same degree of freedom claim for the duality-invariant $s\geq 3$ equations 
requires further analysis of tertiary and higher order constraints. 
 
\section{Discussion}

We have demonstrated
that there are many non-linear generalizations of the $s=2$ PM equations of motion (and possibly also for $s\geq 3$). It is unlikely that these enjoy 
a covariant, local Lagrangian description
since vertices for PM interactions are subject to various no-go results~(see~\cite{Joung} and references therein).
However, non-Lagrangian theories are still potentially of physical interest, especially 
if they enjoy additional symmetries. Since the equations we write are covariant, the models enjoy dS isometries as symmetries.  Moreover, we have identified examples  that also exhibit a duality invariance.

Concerning the existence of action principles for our models, consider  free $s=2$ 
PM. The generating functional for the constitutive equations is then 
$$
S(F)=-\frac14 \int d^4x \sqrt{-g}\, F_{\mu\nu\alpha} F^{\mu\nu\alpha}\, .
$$
For the electromagnetic models, the generating functional $S(F)=-\frac14 \int d^4x \sqrt{-g}\, F_{\mu\nu} F^{\mu\nu}$ also defines the theory's  action upon setting $F=dA$, but for  the $s=2$ PM model this is no longer the case\footnote{
The PM solutions are only a subset of the extrema of this functional:  
set $F(h)_{\mu\nu\alpha}=\nabla_{\mu} h_{\nu\alpha}-\nabla_{\nu} h_{\mu\alpha}$ where the rank~2 tensor $h$ has no definite symmetry. This gives  equations of motion $\nabla^\mu F(h)_{\mu\nu\alpha}=0$ which yield
the PM equations upon truncating $h$ to its symmetric part.}. This general feature of all the models we have presented  may preclude the existence of covariant action principles.

\appendix

\section{Additional spin~2 models?}\label{6manifold}

Spin~2 constitutive relations $G(F)_{\mu\nu\alpha}$ generated by functionals $S(F)$ depending only on curvature bilinears  are interesting because  one can then independently
perform  the~$3+1$ decomposition of~\cite{Hintman}  for both $F_{\mu\nu\alpha}$ and $G(F)_{\mu\nu\alpha}$. The constitutive relation then respects this~$3+1$ split by explicitly relating the spin 2 analogs of the electric intensity and magnetic induction to the electric and magnetic fields.

Independent $s=2$ curvature bilinears  are  given by\footnote{The remaining bilinears obey
$$
F_{\widetilde{\mu\nu}}{}^\nu F^{\widetilde{\mu\rho}}{}_\rho=
-\frac12 \alpha+\gamma\, ,\ 
F_{\widetilde{\mu\nu}\rho}F^{\nu\widetilde{\mu\rho}}=
\frac12\gamma
-\frac12 \epsilon\, ,\ 
\frac12 \epsilon^{\mu\nu\alpha\beta}F_{\mu\rho\nu}F_{\alpha}{}^\rho{}_{\beta}
=\frac12 \delta -\frac 12 \eta\, ,
$$
$$
\frac12 F_{\widetilde{\mu\rho}\nu} F^{\widetilde{\alpha\rho}\beta}
\varepsilon_{\alpha\beta}{}^{\mu\nu}=\frac12 \eta - \frac12 \delta\, ,\ 
F_{\widetilde{\mu\nu}\rho}F^{\widetilde{\mu\rho}\nu}=
-\frac12 \alpha +\varepsilon\, ,\ 
F^\nu{}_{\widetilde{\mu\nu}}F^{\mu\rho}{}_\rho=\eta\, ,\ 
F^\nu{}_{\widetilde{\mu\nu}}F^{\widetilde{\mu\rho}}{}_\rho=-\frac12\alpha+\gamma\, .
$$
}
$$
\alpha:=-F_{\mu\nu\rho}F^{\mu\nu\rho}\, ,\ 
\beta:=F_{\widetilde{\mu\nu}\rho} F^{\mu\nu\rho}\, ,\ 
\gamma:= F_{\mu\nu\rho}F^{\mu\rho\nu}\, ,\ 
$$
$$
\delta:= F_{\widetilde{\mu\nu}\rho} F^{\mu\rho\nu}\, ,\ 
\varepsilon:= F_{\mu\nu}{}^\nu F^{\mu\rho}{}_\rho\, ,\ 
\eta:= F_{\widetilde{\mu\nu}}{}^\nu F^{\mu\rho}{}_\rho\, .
$$
The constitutive relations $G(F)_{\mu\nu\alpha}$ stemming from generating functionals $S(F)=S(\alpha, \beta, \gamma, \delta, \epsilon, \eta)$ are then
$$
G(F)_{\mu\nu\rho}=-4S_\alpha F_{\mu\nu\rho}
-4S_\beta F_{\widetilde{\mu\nu}\rho}
-4 S_{\gamma} F_{[\mu|\rho|\nu]}
-2 S_\delta (F_{\widetilde{\mu\rho}\nu}-F_{\widetilde{\nu\rho}\mu})
-4 S_\varepsilon F_{[\mu} g_{\nu]\rho}
-4 S_\eta F_{\widetilde{[\mu|\sigma}}{}^\sigma g_{\nu]\rho}\, ,
$$
where $S_A:=\frac{\partial S}{\partial x^A}$ for $x^A\in \{\alpha,\beta,\gamma,\delta,\varepsilon,\eta\}$.
Thus the involutive requirement \eqref{EBDHs} becomes
\begin{eqnarray*}
\beta/16&=&\alpha\,  (-2 S_\alpha S_\beta -S_\alpha S_\delta -S_\alpha S_\eta +S_\gamma S_\eta) \\
&&\quad+\, \beta\,  (S_\alpha^2-S_\beta^2)\\
&&\quad+\, \gamma\, (2S_\alpha S_\eta-2S_\beta S_\gamma-S_\gamma S_\delta-2S_\gamma S_\eta)\\
&&\quad+\, \delta\, (\frac12 S_\gamma^2-\frac12 S_\delta^2 +2 S_\alpha S_\gamma -2 S_\beta S_\delta)\\
&&\quad+\, \varepsilon\, (2S_\alpha S_\delta -2 S_\beta S_\varepsilon + S_\gamma S_\delta + 2 S_\delta S_\varepsilon)\\
&&\quad+\, \eta\, (-\frac12 S_\gamma^2+\frac12 S_\delta^2+ 2S_\alpha S_\varepsilon -2 S_\beta S_\eta -2 S_\gamma S_\varepsilon + 2 S_\delta S_\eta)\, ,
\end{eqnarray*}
which determines the (inverse) metric $G^{AB}$ on the 6-dimensional Riemannian manifold coordinatized by the independent bilinears according to 
$$
G^{AB}S_{A}S_{B}=1\, .
$$
This is a unit vector problem 
whose solutions determine generating functions 
for duality invariant models. One such solution is given in~\eqref{bigformaggio}.

\section*{Acknowledgements}
We thank K. Hinterbichler  for useful  discussions. S.D. was supported in part by grants NSF PHY-1266107 and DOE \# de-sc0011632. A.W.
was supported in part by a Simons Foundation Collaboration Grant for Mathematicians. G.Z. was supported in part by DOE Grant DE-FG03-91ER40674.

\end{document}